# All-Optical Information Processing Capacity of Diffractive Surfaces

Onur Kulce[1,2,3,§], Deniz Mengu[1,2,3,§], Yair Rivenson[1,2,3], Aydogan Ozcan[1,2,3,*]

[1] Electrical and Computer Engineering Department, University of California, Los Angeles, CA, 90095, USA

[2] Bioengineering Department, University of California, Los Angeles, CA, 90095, USA

[3] California NanoSystems Institute, University of California, Los Angeles, CA, 90095, USA

[§] Equal contribution

* Corresponding author: ozcan@ucla.edu

*Onur Kulce*: onurkulce@ucla.edu

*Deniz Mengu*: denizmengu@ucla.edu

*Yair Rivenson*: rivensonyair@ucla.edu

*Aydogan Ozcan*: ozcan@ucla.edu

## Abstract

The precise engineering of materials and surfaces has been at the heart of some of the recent advances in optics and photonics. These advances related to the engineering of materials with new functionalities have also opened up exciting avenues for designing trainable surfaces that can perform computation and machine learning tasks through light-matter interactions and diffraction. Here, we analyse the information processing capacity of coherent optical networks formed by diffractive surfaces that are trained to perform an all-optical computational task between a given input and output field-of-view. We show that the dimensionality of the all-optical solution space covering the complex-valued transformations between the input and output fields-of-view is linearly proportional to the number of diffractive surfaces within the optical network, up to a limit that is dictated by the extent of the input and output fields-of-view. Deeper diffractive networks that are composed of larger numbers of trainable surfaces can cover a higher-dimensional subspace of the complex-valued linear transformations between a larger input field-of-view and a larger output field-of-view and exhibit depth advantages in terms of their statistical inference, learning and generalization capabilities for different image classification tasks when compared with a single trainable diffractive surface. These analyses and conclusions are broadly applicable to various forms of diffractive surfaces, including, e.g., plasmonic and/or dielectric-based metasurfaces and flat optics, which can be used to form all-optical processors.

## 1. Introduction

The ever-growing area of engineered materials has empowered the design of novel components and devices that can interact with and harness electromagnetic waves in unprecedented and unique ways, offering various new functionalities [1–14]. Owing to the precise control of material structure and properties as well as the associated light-matter interaction at different scales, these engineered material systems, including, e.g., plasmonics, metamaterials/metasurfaces and flat optics, have led to fundamentally new capabilities in the imaging and sensing fields, among others [15–24]. Optical computing and information processing constitute yet another area that has harnessed engineered light-matter interactions to perform computational tasks using wave optics and the propagation of light through specially devised materials[25–38]. These approaches and many others highlight the emerging uses of trained materials and surfaces as the workhorse of optical computation.

Here, we investigate the information processing capacity of trainable diffractive surfaces to shed light on their computational power and limits. An all-optical diffractive network is physically formed by a number of diffractive layers/surfaces and the free-space propagation between them (see Fig. 1a). Individual transmission and/or reflection coefficients (i.e., neurons) of diffractive surfaces are adjusted or trained to perform a desired input-output transformation task as the light diffracts through these layers. Trained with deep-learning-based error back-propagation methods, these diffractive networks have been shown to perform machine learning tasks such as image classification and deterministic optical tasks including, e.g., wavelength demultiplexing, pulse shaping and imaging[38–44].

The forward model of a diffractive optical network can be mathematically formulated as a complex-valued matrix operator that multiplies an input field vector to create an output field vector at the detector plane/aperture. This operator is designed/trained using, e.g., deep learning to transform a set of complex fields (forming, e.g., the input data classes) at the input aperture of the optical network into another set of corresponding fields at the output aperture (forming, e.g., the data classification signals) and is physically created through the interaction of the input light with the designed diffractive surfaces as well as free-space propagation within the network (Fig. 1a).

In this paper, we investigate the dimensionality of the all-optical solution space that is covered by a diffractive network design as a function of the number of diffractive surfaces, the number of neurons per surface, and the size of the input and output fields-of-view. With our theoretical and numerical analysis, we show that the dimensionality of the transformation solution space that can be accessed through the task-specific design of a diffractive network is linearly proportional to the number of diffractive surfaces, up to a limit that is governed by the extent of the input and output fields-of-view. Stated differently, adding new diffractive surfaces into a given network design increases the dimensionality of the solution space that can be all-optically processed by the diffractive network until it reaches the linear transformation capacity dictated by the input and output apertures (Fig. 1a). Beyond this limit, the addition of new trainable diffractive surfaces into the optical network can cover a higher-dimensional solution space over larger input and output fields-of-view, extending the space-bandwidth product of the all-optical processor.

Our theoretical analysis further reveals that, in addition to increasing the number of diffractive surfaces within a network, another strategy to increase the all-optical processing capacity of a diffractive network is to increase the number of trainable neurons per diffractive surface. However, our numerical analysis involving different image classification tasks demonstrates that this strategy of creating a higher-numerical-aperture (NA) optical network for all-optical processing of the input information is not as effective as increasing the number of diffractive surfaces in terms of the blind inference and generalization performance of the network. Overall, our theoretical and numerical analyses support each other, revealing that deeper diffractive networks with larger numbers of trainable diffractive surfaces exhibit depth advantages in terms of their statistical inference and learning capabilities compared with a single trainable diffractive surface.

The presented analyses and conclusions are generally applicable to the design and investigation of various coherent all-optical processors formed by diffractive surfaces such as, e.g., metamaterials, plasmonic or dielectric-based metasurfaces, and flat-optics-based designer surfaces that can form information processing networks to execute a desired computational task between an input and output aperture.

## 2. Results

### 2.1. Theoretical Analysis of the Information Processing Capacity of Diffractive Surfaces

Let the $\boldsymbol{x}$ and $\boldsymbol{y}$ vectors represent the sampled optical fields (including the phase and amplitude information) at the input and output apertures, respectively. We assume that the sizes of $\boldsymbol{x}$ and $\boldsymbol{y}$ are $N_i \times 1$ and $N_o \times 1$, defined by the input and output fields-of-view, respectively (see Fig. 1a); these two quantities, $N_i$ and $N_o$, are simply proportional to the space-bandwidth product of the input field and the output field at the input and output apertures of the diffractive network, respectively. Outside the input field-of-view (FOV) defined by $N_i$, the rest of the points within the input plane do not transmit light or any information to the diffractive network, i.e., they are assumed to be blocked by, for example, an aperture. In a diffractive optical network composed of transmissive and/or reflective surfaces that rely on linear optical materials, these vectors are related to each other by $\boldsymbol{Ax} = \boldsymbol{y}$, where $\boldsymbol{A}$ represents the combined effects of the free-space wave propagation and the transmission through (or reflection off of) the diffractive surfaces, where the size of $\boldsymbol{A}$ is $N_o \times N_i$. The matrix $\boldsymbol{A}$ can be considered the mathematical operator that represents the all-optical processing of the information carried by the input complex field (within the input field-of-view/aperture), delivering the processing results to the desired output field-of-view.

Here, we prove that an optical network having a larger number of diffractive surfaces or trainable neurons can generate a richer set for the transformation matrix $\boldsymbol{A}$ up to a certain limit within the set of all complex-valued matrices with size $N_o \times N_i$. Therefore, this section analytically investigates the all-optical information processing capacity of diffractive networks composed of diffractive surfaces. The input field is assumed to be monochromatic, spatially and temporally coherent with an arbitrary polarization state, and the diffractive surfaces are assumed to be linear, without any coupling to other states of polarization, which is ignored.

Let $\boldsymbol{H_d}$ be an $N \times N$ matrix, which represents the Rayleigh-Sommerfeld diffraction between two fields specified over parallel planes that are axially separated by a distance $d$. Since $\boldsymbol{H_d}$ is created from the free-space propagation convolution kernel, it is a Toeplitz matrix. Throughout the paper, without loss of generality, we assume that $N_i = N_o = N_{FOV}$, $N \geq N_{FOV}$ and that the diffractive surfaces are separated by free space, i.e., the refractive index surrounding the diffractive layers is taken as $n$ = 1. We also assume that the optical fields include only the propagating modes, i.e., travelling waves; stated differently, the evanescent modes along the propagation direction are not included in our model since $d \geq \lambda$ (Fig. 1b). With this assumption, we choose the sampling period of the discretized complex fields to be $\lambda/2$, where $\lambda$ is the wavelength of the monochromatic input field. Accordingly, the eigenvalues of $\boldsymbol{H_d}$ are in the form $e^{jk_z d}$ for $0 \leq k_z \leq k_o$, where $k_o$ is the wavenumber of the optical field[45].

Furthermore, let $\boldsymbol{T_k}$ be an $N_{Lk} \times N_{Lk}$ matrix, which represents the $k^{\text{th}}$ diffractive surface/layer in the network model, where $N_{Lk}$ is the number of neurons in the corresponding diffractive surface; for a diffractive network composed of $K$ surfaces, without loss of generality we assume $\min(N_{L1}, N_{L2}, \dots, N_{LK}) \geq N_{FOV}$. Based on these definitions, the elements of $\boldsymbol{T_k}$ are nonzero *only* along its main diagonal entries. These diagonal entries represent the complex-valued transmittance (or reflectance) values (i.e., the optical neurons) of the associated diffractive surface, with a sampling period of $\lambda/2$. Furthermore, each diffractive surface defined by a given transmittance matrix is assumed to be surrounded by a blocking layer within the same plane to avoid any optical communication between the layers without passing through an intermediate diffractive surface. This formalism embraces any form of diffractive surface, including, e.g., plasmonic or dielectric-based metasurfaces. Even if the diffractive surface has deeply sub-wavelength structures, with a much smaller sampling period compared to $\lambda/2$ and many more degrees of freedom (*M*) compared to $N_{Lk}$, the information processing capability of a diffractive surface within a network is limited to propagating modes since $d \geq \lambda$, which restricts the effective number of neurons per layer to $N_{Lk}$ (Fig. 1b). In other words, since we assume that only propagating modes can reach the subsequent diffractive surfaces within the optical diffractive network, the sampling period (and hence, the neuron size) of $\lambda/2$ is sufficient to represent these propagating modes in air[46]. According to Shannon's sampling theorem, since the spatial frequency band of the propagating modes in air is restricted to the $(-1/\lambda, 1/\lambda)$ interval, a neuron size that is smaller than $\lambda/2$ leads to oversampling and over-utilization of the optical neurons of a given diffractive surface. On the other hand, if one aims to control and engineer the evanescent modes, then a denser sampling period on each diffractive surface is needed, which might be useful to build diffractive networks that have $d \ll \lambda$. In this near-field diffractive network, the enormously rich degrees of freedom enabled by various metasurface designs with $M \gg N_{Lk}$ can be utilized to provide full and independent control of the phase and amplitude coefficients of each individual neuron of a diffractive surface.

The underlying physical process of how the light is modulated by an optical neuron may vary in different diffractive surface designs. In a dielectric-material-based transmissive design, for example, phase modulation can be achieved by slowing down the light inside the material, where the thickness of an optical neuron determines the amount of phase shift that the light beam undergoes. Alternatively, liquid-crystal (LC)-based spatial light modulators (SLMs) or flat-optics-based metasurfaces can also be employed as part of a diffractive network to generate the desired phase and/or amplitude modulation on the transmitted or reflected light[9,47].

Starting from Section 2.1.1, we investigate the physical properties of $\boldsymbol{A}$, generated by different numbers of diffractive surfaces and trainable neurons. In this analysis, without loss of generality, each diffractive surface is assumed to be transmissive, following the schematics shown in Fig. 1a, and its extension to reflective surfaces is straightforward and does not change our conclusions. Finally, multiple (back and forth) reflections within a diffractive network composed of different layers are ignored in our analysis, as these are much weaker processes compared to the forward propagating modes.

### 2.1.1. Analysis of a single diffractive surface

The input-output relationship for a single diffractive surface that is placed between an input and an output FOV (Fig. 1a) can be written as:

$$\boldsymbol{y} = \boldsymbol{H}'_{\boldsymbol{d_2}} \boldsymbol{T_1} \boldsymbol{H}'_{\boldsymbol{d_1}} \boldsymbol{x} = \boldsymbol{A_1} \boldsymbol{x} \quad (1)$$

where $d_1 \geq \lambda$ and $d_2 \geq \lambda$ represent the axial distance between the input plane and the diffractive surface, and the axial distance between the diffractive surface and the output plane, respectively. Here we also assume that $d_1 \neq d_2$; the Supplementary Information, Section S5 discusses the special case of $d_1 = d_2$. Since there is only one diffractive surface in the network, we denote the transmittance matrix as $\boldsymbol{T_1}$, the size of which is $N_{L1} \times N_{L1}$, where $L1$ represents the diffractive surface. Here, $\boldsymbol{H}'_{\boldsymbol{d_1}}$is an $N_{L1} \times N_{FOV}$ matrix that is generated from the $N_{L1} \times N_{L1}$ propagation matrix $\boldsymbol{H_{d_1}}$ by deleting the appropriately chosen $N_{L1} - N_{FOV}$-many columns. The positions of the deleted columns correspond to the zero transmission values at the input plane that lie outside the input field-of-view or aperture defined by $N_i = N_{FOV}$ (Fig. 1a), i.e., not included in $\boldsymbol{x}$. Similarly, $\boldsymbol{H}'_{\boldsymbol{d_2}}$ is an $N_{FOV} \times N_{L1}$ matrix that is generated from the $N_{L1} \times N_{L1}$ propagation matrix $\boldsymbol{H_{d_2}}$ by deleting the appropriately chosen $N_{L1} - N_{FOV}$-many rows, which correspond to the locations outside the output FOV or aperture defined by $N_o = N_{FOV}$ in Fig. 1a; this means that the output field is calculated only within the desired output aperture. As a result, $\boldsymbol{H}'_{\boldsymbol{d_1}}$ and $\boldsymbol{H}'_{\boldsymbol{d_2}}$ have a rank of $N_{FOV}$.

To investigate the information processing capacity of $\boldsymbol{A_1}$ based on a single diffractive surface, we vectorize this matrix in the column order and denote it as $vec(\boldsymbol{A_1}) = \boldsymbol{a_1}$ [48]. Next, we show that the set of possible $\boldsymbol{a_1}$ vectors forms a $\min(N_{L1}, N_{FOV}^2)$-dimensional subset of an $N_{FOV}^2$-dimensional complex-valued vector field. The vector, $\boldsymbol{a_1}$, can be written as:

$$\begin{aligned} vec(\boldsymbol{A_1}) = \boldsymbol{a_1} &= vec\left(\boldsymbol{H}'_{\boldsymbol{d_2}} \boldsymbol{T_1} \boldsymbol{H}'_{d_1}\right) \\ &= \left(\boldsymbol{H}'^{T}_{\boldsymbol{d_1}} \otimes \boldsymbol{H}'_{\boldsymbol{d_2}}\right) vec(\boldsymbol{T_1}) \\ &= \left(\boldsymbol{H}'^{T}_{\boldsymbol{d_1}} \otimes \boldsymbol{H}'_{\boldsymbol{d_2}}\right) \boldsymbol{t_1} \end{aligned} \quad (2)$$

where the superscript $T$ and $\otimes$ denote the transpose operation and Kronecker product, respectively[48]. Here, the size of $\boldsymbol{H}'^{T}_{\boldsymbol{d_1}} \otimes \boldsymbol{H}'_{\boldsymbol{d_2}}$ is $N_{FOV}^2 \times N_{L1}^2$, and it is a full-rank matrix with rank $N_{FOV}^2$. In Equation 2, $vec(\boldsymbol{T_1}) = \boldsymbol{t_1}$ has at most $N_{L1}$ controllable/adjustable complex-valued entries, which physically represent the neurons of the diffractive surface, and the rest of its entries are all zero. These transmission coefficients lead to a linear combination of $N_{L1}$-many vectors of $\boldsymbol{H}'^{T}_{\boldsymbol{d_1}} \otimes \boldsymbol{H}'_{\boldsymbol{d_2}}$, where $d_1 \neq d_2 \neq 0$. If $N_{L1} \leq N_{FOV}^2$, these vectors subject to the linear combination are linearly independent (see the Supplementary Information Section S4.1 and Supplementary Figure S1). Hence, the set of resulting $\boldsymbol{a_1}$ vectors generated by Equation 2 forms an $N_{L1}$-dimensional subspace of the $N_{FOV}^2$-dimensional complex-valued vector space. On the other hand, if $N_{L1} > N_{FOV}^2$, then the vectors in the linear combination start to become dependent on each other. In this case of $N_{L1} > N_{FOV}^2$, the dimensionality of the set of possible vector fields is limited to $N_{FOV}^2$ (also see Supplementary Figure S1).

This analysis demonstrates that the set of complex field transformation vectors that can be generated by a single diffractive surface that connects a given input and output FOV constitutes a $\min(N_{L1}, N_{FOV}^2)$-dimensional subspace of an $N_{FOV}^2$-dimensional complex-valued vector space. These results are based on our earlier assumption that $d_1 \geq \lambda$, $d_2 \geq \lambda$ and $d_1 \neq d_2$. For the special case of $d_1 = d_2 \geq \lambda$, the upper limit of the dimensionality of the solution space that can be generated by a single diffractive surface ($K$ = 1) is reduced from $N_{FOV}^2$ to $(N_{FOV}^2 + N_{FOV})/2$ due to the combinatorial symmetries that exist in the optical path for $d_1 = d_2$ (see the Supplementary Information, Section S5).

### 2.1.2. Analysis of an optical network formed by two diffractive surfaces

Here, we consider an optical network with two different (trainable) diffractive surfaces ($K$=2), where the input-output relation can be written as:

$$\boldsymbol{y} = \boldsymbol{H'_{d_3} T_2 H_{d_2} T_1 H'_{d_1} x} = \boldsymbol{A_2 x} \tag{3}$$

$N_x = \max(N_{L1}, N_{L2})$ determines the sizes of the matrices in Equation 3, where $N_{L1}$ and $N_{L2}$ represent the number of neurons in the first and second diffractive surfaces, respectively; $d_1$, $d_2$ and $d_3$ represent the axial distances between the diffractive surfaces (see Fig. 1a). Accordingly, the sizes of $\boldsymbol{H'_{d_1}}$, $\boldsymbol{H_{d_2}}$ and $\boldsymbol{H'_{d_3}}$ become $N_x \times N_{FOV}$, $N_x \times N_x$ and $N_{FOV} \times N_x$, respectively. Since we have already assumed that $\min(N_{L1}, N_{L2}) \geq N_{FOV}$, $\boldsymbol{H'_{d_1}}$ and $\boldsymbol{H'_{d_3}}$ can be generated from the corresponding $N_x \times N_x$ propagation matrices by deleting the appropriate columns and rows, as described in Section 2.1.1. Because $\boldsymbol{H_{d_2}}$ has a size of $N_x \times N_x$, there is no need to delete any rows or columns from the associated propagation matrix. Although both $\boldsymbol{T_1}$ and $\boldsymbol{T_2}$ have a size of $N_x \times N_x$, the one corresponding to the diffractive surface that contains the smaller number of neurons has some zero values along its main diagonal indices. The number of these zeros is $N_x - \min(N_{L1}, N_{L2})$.

Similar to the analysis reported in Section 2.1.1, the vectorization of $\boldsymbol{A_2}$ reveals:

$$\begin{aligned} vec(\boldsymbol{A_2}) = \boldsymbol{a_2} &= vec\left(\boldsymbol{H'_{d_3} T_2 H_{d_2} T_1 H'_{d_1}}\right) \\ &= \left(\boldsymbol{H'^{T}_{d_1}} \otimes \boldsymbol{H'_{d_3}}\right) vec\left(\boldsymbol{T_2 H_{d_2} T_1}\right) \\ &= \left(\boldsymbol{H'^{T}_{d_1}} \otimes \boldsymbol{H'_{d_3}}\right)(\boldsymbol{T_1^T} \otimes \boldsymbol{T_2}) vec\left(\boldsymbol{H_{d_2}}\right) \\ &= \left(\boldsymbol{H'^{T}_{d_1}} \otimes \boldsymbol{H'_{d_3}}\right)(\boldsymbol{T_1} \otimes \boldsymbol{T_2}) vec\left(\boldsymbol{H_{d_2}}\right) \\ &= \left(\boldsymbol{H'^{T}_{d_1}} \otimes \boldsymbol{H'_{d_3}}\right)(\boldsymbol{T_1} \otimes \boldsymbol{T_2}) \boldsymbol{h_{d_2}} \\ &= \left(\boldsymbol{H'^{T}_{d_1}} \otimes \boldsymbol{H'_{d_3}}\right) \boldsymbol{\widehat{H}_{d_2}} diag(\boldsymbol{T_1} \otimes \boldsymbol{T_2}) \\ &= \left(\boldsymbol{H'^{T}_{d_1}} \otimes \boldsymbol{H'_{d_3}}\right) \boldsymbol{\widehat{H}_{d_2} t_{12}} \end{aligned} \tag{4}$$

where $\widehat{\boldsymbol{H}}_{\boldsymbol{d_2}}$ is an $N_x^2 \times N_x^2$ matrix that has nonzero entries *only* along its main diagonal locations. These entries are generated from $vec(\boldsymbol{H}_{\boldsymbol{d_2}}) = \boldsymbol{h}_{\boldsymbol{d_2}}$ such that $\widehat{\boldsymbol{H}}_{\boldsymbol{d_2}}[i,i] = \boldsymbol{h}_{\boldsymbol{d_2}}[i]$. Since the $diag(\cdot)$ operator forms a vector from the main diagonal entries of its input matrix, the vector $\boldsymbol{t_{12}} = diag(\boldsymbol{T_1} \otimes \boldsymbol{T_2})$ is generated such that $\boldsymbol{t_{12}}[i] = (\boldsymbol{T_1} \otimes \boldsymbol{T_2})[i,i]$. The equality $(\boldsymbol{T_1} \otimes \boldsymbol{T_2})\boldsymbol{h}_{\boldsymbol{d_2}} = \widehat{\boldsymbol{H}}_{\boldsymbol{d_2}}\boldsymbol{t_{12}}$ stems from the fact that the nonzero elements of $\boldsymbol{T_1} \otimes \boldsymbol{T_2}$ are located only along its main diagonal entries.

In Equation 4, $\boldsymbol{H}'^T_{\boldsymbol{d_1}} \otimes \boldsymbol{H}'_{\boldsymbol{d_3}}$ has rank $N_{FOV}^2$. Since all the diagonal elements of $\widehat{\boldsymbol{H}}_{\boldsymbol{d_2}}$are nonzero, it has rank $N_x^2$. As a result, $(\boldsymbol{H}^T_{\boldsymbol{d_1}} \otimes \boldsymbol{H}_{\boldsymbol{d_3}})\widehat{\boldsymbol{H}}_{\boldsymbol{d_2}}$ is a full-rank matrix with rank $N_{FOV}^2$. Additionally, the nonzero elements of $\boldsymbol{t_{12}}$ take the form $t_{ij} = t_{1,i}t_{2,j}$, where $t_{1,i}$ and $t_{2,j}$ are the trainable/adjustable complex transmittance values of the $i^{\text{th}}$ neuron of the 1st diffractive surface and the $j^{\text{th}}$ neuron of the 2nd diffractive surface, respectively, for $i \in \{1,2,\dots,N_{L1}\}$ and $j \in \{1,2,\dots,N_{L2}\}$. Then, the set of possible $\boldsymbol{a_2}$ vectors (Equation 4) can be written as:

$$\boldsymbol{a_2} = \sum_{i,j} t_{ij}\boldsymbol{h_{ij}} \qquad (5)$$

where $\boldsymbol{h_{ij}}$ is the corresponding column vector of $(\boldsymbol{H}'^T_{\boldsymbol{d_1}} \otimes \boldsymbol{H}'_{\boldsymbol{d_3}})\widehat{\boldsymbol{H}}_{\boldsymbol{d_2}}$.

Equation 5 is in the form of a complex-valued linear combination of $N_{L1}N_{L2}$-many complex-valued vectors, $\boldsymbol{h_{ij}}$. Since we assume $\min(N_{L1}, N_{L2}) \geq N_{FOV}$, these vectors necessarily form a linearly dependent set of vectors and this restricts the dimensionality of the vector space to $N_{FOV}^2$. Moreover, due to the coupling of the complex-valued transmittance values of the two diffractive surfaces ($t_{ij} = t_{1,i}t_{2,j}$) in Equation 5, the dimensionality of the resulting set of $\boldsymbol{a_2}$ vectors can even go below $N_{FOV}^2$, despite $N_{L1}N_{L2} \geq N_{FOV}^2$. In fact, in the Materials and Methods section, we show that the set of $\boldsymbol{a_2}$ vectors can form an $N_{L1}+N_{L2}-1$-dimensional subspace of the $N_{FOV}^2$-dimensional complex-valued vector space and can be written as:

$$\boldsymbol{a_2} = \sum_{k=1}^{N_{L1}+N_{L2}-1} c_k\boldsymbol{b_k} \qquad (6)$$

where $\boldsymbol{b_k}$ represents length-$N_{FOV}^2$ linearly independent vectors and $c_k$ represents complex-valued coefficients, generated through the coupling of the transmittance values of the two independent diffractive surfaces. The relationship between Equations 5 and 6 is also presented as a pseudo-code in Table 1; see also Supplementary Tables S1-S3 and Supplementary Figure S2.

These analyses reveal that by using a diffractive optical network composed of two different trainable diffractive surfaces (with neurons $N_{L1}, N_{L2}$), it is possible to generate an all-optical solution that spans an $N_{L1}+N_{L2}-1$ dimensional subspace of an $N_{FOV}^2$-dimensional complex-valued vector space. As a special case, if we assume $N = N_{L1} = N_{L2} = N_i = N_o = N_{FOV}$, the resulting set of complex-valued linear transformation vectors forms a $2N-1$ dimensional subspace of an $N^2$-dimensional vector field. The Supplementary Information (Section S1 and Table S1) also provides a coefficient and basis vector generation algorithm, independently reaching the same conclusion that this special case forms a $2N-1$ dimensional subspace of an $N^2$-dimensional vector field. The upper limit of the solution space dimensionality that can be achieved by a two-layered diffractive network is $N_{FOV}^2$, which is dictated by the input and output fields-of-view between which the diffractive network is positioned.

In summary, these analyses show that the dimensionality of the all-optical solution space covered by two trainable diffractive surfaces ($K$ = 2) positioned between a given set of input-output FOV is given by $\min(N_{FOV}^2, \mathrm{N}_{L1}+\mathrm{N}_{L2}-1)$. Different from $K$ = 1 architecture, which revealed a restricted solution space when $d_1 = d_2$ (see the Supplementary Information, Section S5), diffractive optical networks with $K$=2 do not exhibit a similar restriction related to the axial distances $d_1$, $d_2$ and $d_3$ (see Supplementary Figure S2).

### 2.1.3. Analysis of an optical network formed by three or more diffractive surfaces

Next, we consider an optical network formed by more than 2 diffractive surfaces, with neurons of $(N_{L1}, N_{L2}, \cdots, N_{LK})$ for each layer, where $K$ is the number of diffractive surfaces and $N_{Lk}$ represents the number of neurons in the $k^{th}$ layer. In the previous section, we showed that a two-layered network with $(N_{L1}, N_{L2})$ neurons has the same solution space dimensionality as that of a single-layered, larger diffractive network having $\mathrm{N}_{L1}+\mathrm{N}_{L2}-1$ individual neurons. If we assume that a third diffractive surface ($N_{L3}$) is added to this single-layer network with $\mathrm{N}_{L1}+\mathrm{N}_{L2}-1$ neurons, this becomes equivalent to a two-layered network with $(N_{L1}+N_{L2}-1\,, N_{L3})$ neurons. Based on Section 2.1.2, the dimensionality of the all-optical solution space covered by this diffractive network positioned between a set of input-output fields-of-view is given by $\min(N_{FOV}^2, N_{L1}+N_{L2}+N_{L3}-2)$; also see Supplementary Figure S3. For the special case of $N_{L1} = N_{L2} = N_{L3} = N_i = N_o = N$, Supplementary Information Section S2 and Table S2 independently illustrate that the resulting vector field is indeed a $3N-2$ dimensional subspace of an $N^2$-dimensional vector field.

The above arguments can be extended to a network that has $K$ diffractive surfaces. That is, for a multi-surface diffractive network with a neuron distribution of $(N_{L1}, N_{L2}, \cdots, N_{LK})$, the dimensionality of the solution space (see Fig. 2) created by this diffractive network is given by:

$$\min\left(N_{FOV}^2, \left[\sum_{k=1}^{K} N_{Lk}\right] - (K-1)\right) \qquad (7)$$

which forms a subspace of an $N_{FOV}^2$-dimensional vector space that covers all the complex-valued linear transformations between the input and output fields-of-view.

The upper bound on the dimensionality of the solution space, i.e., the $N_{FOV}^2$ term in Equation 7, is heuristically imposed by the number of possible ray interactions between the input and output fields-of-view. That is, if we consider the diffractive optical network as a black box (Fig. 1a), its operation can be intuitively understood as controlling the phase and/or amplitude of the light rays that are collected from the input, to be guided to the output, following a lateral grid of $\lambda/2$ at the input/output fields-of-view, determined by the diffraction limit of light. The second term in Equation 7, on the other hand, reflects the total space-bandwidth product of $K$ successive diffractive surfaces, one following another. To intuitively understand the $(K-1)$ subtraction term in Equation 7, one can hypothetically consider the simple case of $N_{Lk}=N_{FOV}=1$ for all $K$ diffractive layers; in this case, $[\sum_{k=1}^{K} N_{Lk}]-(K-1)=1$, which simply indicates that $K$ successive diffractive surfaces (each with $N_{Lk}=1$) are equivalent, as physically expected, to a single controllable diffractive surface with $N_L$=1.

Without loss of generality, if we assume $N=N_k$ for all the diffractive surfaces, then the dimensionality of the linear transformation solution space created by this diffractive network will be $KN-(K-1)$, provided that $KN-(K-1)\leq N_{FOV}^2$. The Supplementary Information (Section S3 and Table S3) also provides an independent proof of the same conclusion. This means that for a fixed design choice of $N$ neurons per diffractive surface (determined by, e.g., the limitations of the fabrication methods or other practical considerations), adding new diffractive surfaces to the same diffractive network linearly increases the dimensionality of the solution space that can be all-optically processed by the diffractive network between the input/output fields-of-view. As we further increase $K$ such that $KN-(K-1)\geq N_{FOV}^2$, the diffractive network reaches its linear transformation capacity, and adding more layers or more neurons to the network does not further contribute to its processing power for the desired input-output fields-of-view (see Fig. 2). However, these deeper diffractive networks that have larger numbers of diffractive surfaces (i.e., $KN-(K-1)\geq N_{FOV}^2$) can cover a solution space with a dimensionality of $KN-(K-1)$ over larger input and output fields-of-view. Stated differently, for any given choice of $N$ neurons per diffractive surface, deeper diffractive networks that are composed of multiple surfaces can cover a $KN-(K-1)$-dimensional subspace of all the complex-valued linear transformations between a larger input field-of-view ($N'_i > N_i$) and/or a larger output field-of view ($N'_o > N_o$), as long as $KN-(K-1)\leq N'_i N'_o$. The conclusions of this analysis are also summarized in Fig. 2.

In addition to increasing $K$ (the number of diffractive surfaces within an optical network), an alternative strategy to increase the all-optical processing capabilities of a diffractive network is to increase $N$, the number of neurons per diffractive surface/layer. However, as we numerically demonstrate in the next section, this strategy is not as effective as increasing the number of diffractive surfaces since deep-learning-based design tools are relatively inefficient in utilizing all the degrees of freedom provided by a diffractive surface with $N >> N_o, N_i$. This is partially related to the fact that high-numerical-aperture optical systems are generally more difficult to optimize and design. Moreover, if we consider a single-layer diffractive network design with a large $N_{\max}$ (which defines the *maximum* surface area that can be fabricated and engineered with the desired transmission coefficients), even for this $N_{\max}$ design, the addition of new diffractive surfaces with $N_{\max}$ at each surface linearly increases the dimensionality of the solution space created by the diffractive network, covering linear transformations over larger input and output fields-of-view, as discussed earlier. These reflect some of the important depth advantages of diffractive optical networks that are formed by multiple diffractive surfaces. The next section further expands on this using a numerical analysis of diffractive optical networks that are designed for image classification.

### 2.2. Numerical Analysis of Diffractive Networks

The previous section showed that the dimensionality of the all-optical solution space covered by $K$ diffractive surfaces, forming an optical network positioned between an input and output field-of-view, is determined by $\min\left(N_{FOV}^2, \left[\sum_{k=1}^{K} N_{Lk}\right] - (K-1)\right)$. However, this mathematical analysis does not shed light on the selection or optimization of the complex transmittance (or reflectance) values of each neuron of a diffractive network that is assigned for a given computational task. Here, we numerically investigate the function approximation power of multiple diffractive surfaces in the ($N$, $K$) space using image classification as a computational goal for the design of each diffractive network. Since $N_{FOV}$ and $N$ are large numbers in practice, an iterative optimization procedure based on error back-propagation and deep learning with a desired loss function was used to design diffractive networks and compare their performances as a function of ($N$, $K$).

For the first image classification task that was used as a test-bed, we formed nine different image data classes, where the input field-of-view (aperture) was randomly divided into nine different groups of pixels, each group defining one image class (Fig. 3a). Images of a given data class can have pixels only within the corresponding group, emitting light at arbitrary intensities towards the diffractive network. The computational task of each diffractive network is to blindly classify the input images from one of these nine different classes using *only nine large-area detectors* at the output field-of-view (Fig. 3b), where the classification decision is made based on the *maximum* of the optical signal collected by these nine detectors, each assigned to one particular image class. For deep-learning-based training of each diffractive network for this image classification task, we employed a cross-entropy loss function (see the Materials and Methods section).

Before we report the results of our analysis using a more standard image classification dataset such as CIFAR-10,[49] we initially selected this image classification problem defined in Fig. 3 as it provides a well-defined linear transformation between the input and output fields-of-view. It also has various implications for designing new imaging systems with unique functionalities that cannot be covered by standard lens design principles.

Based on the diffractive network configuration and the image classification problem depicted in Fig. 3, we compared the training and blind testing accuracies provided by different diffractive networks composed of 1, 2 and 3 diffractive surfaces (each surface having $N$ = 40K = 200×200 neurons) under different training and testing conditions (see Figs. 4-5). Our analysis also included the performance of a wider single-layer diffractive network with $N$ = 122.5K > 3×40K neurons. For the training of these diffractive systems, we created two different training image sets ($Tr_1$ and $Tr_2$) to test the learning capabilities of different network architectures. In the first case, the training samples were selected such that approximately 1% of the point sources defining each image data class were simultaneously on and emitting light at various power levels. For this training set, 200K images were created, forming $Tr_1$. In the second case, the training image dataset was constructed to include *only* a single point source (per image) located at different coordinates representing different data classes inside the input field-of-view, providing us with a total of 6.4K training images (which formed $Tr_2$). For the quantification of the blind testing accuracies of the trained diffractive models, three different test image datasets (never used during the training) were created, with each dataset containing 100K images. These three distinct test datasets (named $Te_1$, $Te_{50}$ and $Te_{90}$) contain image samples that take contributions from 1% ($Te_1$), 50% ($Te_{50}$) and 90% ($Te_{90}$) of the points defining each image data class (see Fig. 3).

Figure 4 illustrates the blind classification accuracies achieved by the different diffractive network models that we trained. We see that as the number of diffractive surfaces in the network increases, the testing accuracies achieved by the final diffractive design improve significantly, meaning that the linear transformation space covered by the diffractive network expands with the addition of new trainable diffractive surfaces, in line with our former theoretical analysis. For instance, while a diffractive image classification network with a single phase-only (complex) modulation surface can achieve 24.48% (27.00%) for the test image set $Te_1$, the three-layer versions of the same architectures attain 85.2% (100.00%) blind testing accuracies, respectively (see Figs. 4a,b). Figure 5 shows the phase-only diffractive layers comprising the 1- and 3-layer diffractive optical networks that are compared in Fig. 4a; Fig. 5 also reports some exemplary test images selected from $Te_1$ and $Te_{50}$, along with the corresponding intensity distributions at the output planes of the diffractive networks. The comparison between two- and three-layer diffractive systems also indicates a similar conclusion for the test image set, $Te_1$. However, as we increase the number of point sources contributing to the test images, e.g., for the case of $Te_{90}$, the blind testing classification accuracies of both the two- and three-layer networks saturate at nearly 100%, indicating that the solution space of the two-layer network already covers the optical transformation required to address this relatively easier image classification problem set by $Te_{90}$.

A direct comparison between the classification accuracies reported in Figs. 4a,c and Figs. 4b,d further reveals that the phase-only modulation constraint relatively limits the approximation power of the diffractive network since it places a restriction on the coefficients of the basis vectors, $\boldsymbol{h_{ij}}$. For example, when a two-layer, phase-only diffractive network is trained with $Tr_1$ and blindly tested with the images of $Te_1$, the training and testing accuracies are obtained as 78.72% and 78.44%, respectively. On the other hand, if the diffractive surfaces of the same network architectures have independent control of the transmission amplitude and phase value of each neuron of a given surface, the same training ($Tr_1$) and testing ($Te_1$) accuracy values increase to 97.68% and 97.39%, respectively.

As discussed in our earlier theoretical analysis, an alternative strategy to increase the all-optical processing capabilities of a diffractive network is to increase $N$, the number of neurons per diffractive surface. We also numerically investigated this scenario by training and testing another diffractive image classifier with a single surface that contains 122.5K neurons, i.e., it has more trainable neurons than the 3-layer diffractive designs reported in Fig. 4. As demonstrated in Fig. 4, although the performance of this larger/wider diffractive surface surpassed that of the previous, narrower/smaller 1-layer designs with 40K trainable neurons, its blind testing accuracy could not match the classification accuracies achieved by a 2-layer (2×40K neurons) network in both the phase-only and complex modulation cases. Despite using more trainable neurons than the 2-layer and 3-layer diffractive designs, the blind inference and generalization performance of this larger/wider diffractive surface is worse than that of the multi-surface diffractive designs. In fact, if we were to further increase the number of neurons in this single diffractive surface (further increasing the effective numerical aperture of the diffractive network), the inference performance gain due to these additional neurons that are farther away from the optical axis will asymptotically go to zero since the corresponding k-vectors of these neurons carry a limited amount of optical power for the desired transformations targeted between the input and output fields-of-view.

Another very important observation that one can make in Figs. 4c,d is that the performance improvements due to the increasing number of diffractive surfaces are much more pronounced for more challenging (i.e., limited) training image datasets, such as $Tr_2$. With a significantly smaller number of training images (6.4K images in $Tr_2$ as opposed to 200K images in $Tr_1$), multi-surface diffractive networks trained with $Tr_2$ successfully generalized to different test image datasets ($Te_1$, $Te_{50}$ and $Te_{90}$) and efficiently learned the image classification problem at hand, whereas the single-surface diffractive networks (including the one with 122.5K trainable neurons per layer) almost entirely failed to generalize; see, e.g., Figs. 4c,d, the blind testing accuracy values for the diffractive models trained with $Tr_2$.

Next, we applied our analysis to a widely used, standard image classification dataset and investigated the performance of diffractive image classification networks comprised of one, three and five diffractive surfaces using the CIFAR-10 image dataset[49]. Unlike the previous image classification dataset (Fig. 3), the samples of CIFAR-10 contain images of physical objects, e.g., airplanes, birds, cats, dogs, etc., and CIFAR-10 has been widely used for quantifying the approximation power associated with various deep neural network architectures. Here, we assume that the CIFAR-10 images are encoded in the phase channel of the input field-of-view that is illuminated with a uniform plane wave. For deep-learning-based training of the diffractive classification networks, we adopted two different loss functions. The first loss function is based on the mean-squared-error (MSE), which essentially formulates the design of the all-optical object classification system as an image transformation/projection problem, and the second one is based on the cross-entropy loss, which is commonly used to solve the multi-class separation problems in the deep learning literature (refer to the Materials and Methods section for details).

The results of our analysis are summarized in Figs. 6a and 6b, which report the average blind inference accuracies along with the corresponding standard deviations observed over the testing of three different diffractive network models trained independently to classify the CIFAR-10 test images using phase-only and complex-valued diffractive surfaces, respectively. The 1-, 3-, and 5-layer phase-only (complex-valued) diffractive network architectures can attain blind classification accuracies of 40.55∓0.10% (41.52∓0.09%), 44.47∓0.14% (45.88∓0.28%) and 45.53∓0.30% (46.84∓0.46%), respectively, when they are trained based on the cross-entropy loss detailed in the Materials and Methods section. On the other hand, with the use of the MSE loss, these classification accuracies are reduced to 16.25∓0.48% (14.92∓0.26%), 29.08∓0.14% (33.52∓0.40%) and 33.67∓0.57% (34.69∓0.11%), respectively. In agreement with the conclusions of our previous results and the presented theoretical analysis, the blind testing accuracies achieved by the all-optical diffractive classifiers improve with increasing number of diffractive layers, $K$, independent of the loss function used and the modulation constraints imposed on the trained surfaces (see Fig. 6).

Different from electronic neural networks, however, diffractive networks are physical machine learning platforms with their own optical hardware; hence, practical design merits such as the signal-to-noise ratio (SNR) and the contrast-to-noise ratio (CNR) should also be considered, as these features can be critical for the success of these networks in various applications. Therefore, in addition to the blind testing accuracies, the performance evaluation and comparison of these all-optical diffractive classification systems involve two additional metrics that are analogous to the SNR and CNR. The first is the *classification efficiency*, which we define as the ratio of the optical signal collected by the target, ground-truth class detector, $I_{gt}$, with respect to the total power collected by all class detectors located at the output plane. The second performance metric refers to the normalized difference between the optical signals measured by the ground-truth/correct detector, $I_{gt}$, and its strongest competitor, $I_{sc}$, i.e., $(I_{gt} - I_{sc}) / I_{gt}$; this optical signal contrast metric is, in general, important since the relative level of detection noise with respect to this difference is critical for translating the accuracies achieved by the numerical forward models to the performance of the physically fabricated diffractive networks. Figure 6 reveals that the improvements observed in the blind testing accuracies as a function of the number of diffractive surfaces also apply to these two important diffractive network performance metrics, resulting from the increased dimensionality of the all-optical solution space of the diffractive network with increasing *K*. For instance, the diffractive network models presented in Fig. 6b, trained with the cross-entropy (or MSE) loss function, provide *classification efficiencies* of 13.72∓0.03% (13.98∓0.12%), 15.10∓0.08% (31.74∓0.41%) and 15.46∓0.08% (34.43∓0.28%) using complex-valued 1-, 3- and 5-layers, respectively. Furthermore, the *optical signal contrast* attained by the same diffractive network designs can be calculated as 10.83∓0.17% (9.25∓0.13%), 13.92∓0.28% (35.23∓1.02%) and 14.88∓0.28% (38.67∓0.13%), respectively. Similar improvements are also observed for the phase-only diffractive optical network models that are reported in Fig. 6a. These results indicate that the increased dimensionality of the solution space with increasing *K* improves the inference capacity as well as the robustness of the diffractive network models by enhancing their optical efficiency and signal contrast.

Apart from the results and analyses reported in this section, the depth advantage of diffractive networks has been empirically shown in the literature for some other applications and datasets, such as, e.g., image classification[38,40] and optical spectral filter design[42].

## 3. Discussion

In a diffractive optical design problem, it is not guaranteed that the diffractive surface profiles will converge to the optimum solution for a given (*N*, *K*) configuration. Furthermore, for most applications of interest, such as image classification, the optimum transformation matrix that the diffractive surfaces need to approximate is unknown; for example, what defines all the images of cats vs. dogs (such as in the CIFAR-10 image dataset) is not known analytically to create a target transformation. Nonetheless, it can be argued that as the dimensionality of the all-optical solution space, and thus the approximation power of the diffractive surfaces, increases, the probability of converging to a solution satisfying the desired design criteria also increases. In other words, even if the optimization of the diffractive surfaces becomes trapped in a local minimum, which is practically always the case, there is a greater chance that this state will be closer to the globally optimal solution(s) for deeper diffractive networks with multiple trainable surfaces.

Although not considered in our analysis thus far, an interesting future direction to investigate is the case where the axial distance between two successive diffractive surfaces is made much smaller than the wavelength of light, i.e., $d \ll \lambda$. In this case, all the evanescent waves and the surface modes of each diffractive layer will need to be carefully taken into account to analyse the all-optical processing capabilities of the resulting diffractive network. This would significantly increase the space-bandwidth product of the optical processor as the effective neuron size per diffractive surface/layer can be deeply sub-wavelength if the near-field is taken into account. Furthermore, due to the presence of near-field coupling between diffractive surfaces/layers, the effective transmission or reflection coefficient of each neuron of a surface will no longer be an independent parameter, as it will depend on the configuration/design of the other surfaces. If all of these near-field related coupling effects are carefully taken into consideration during the design of a diffractive optical network with $d \ll \lambda$, it can significantly enrich the solution space of multi-layer coherent optical processors, assuming that the surface fabrication resolution and the signal-to-noise ratio as well as the dynamic range at the detector plane are all sufficient. Despite the theoretical richness of near-field-based diffractive optical networks, the design and implementation of these systems bring substantial challenges in terms of their 3D fabrication and alignment as well as the accuracy of the computational modelling of the associated physics within the diffractive network, including multiple reflections and boundary conditions. While various electromagnetic wave solvers can handle the numerical analysis of near-field diffractive systems, practical aspects of a fabricated near-field diffractive neural network will present various sources of imperfections and errors that might force the physical forward model to significantly deviate from the numerical simulations.

In summary, we presented a theoretical and numerical analysis of the information processing capacity and function approximation power of diffractive surfaces that can compute a given task using temporally and spatially coherent light. In our analysis, we assumed that the polarization state of the propagating light is preserved by the optical modulation on the diffractive surfaces and that the axial distance between successive layers is kept large enough to ensure that the near-field coupling and related effects can be ignored in the optical forward model. Based on these assumptions, our analysis shows that the dimensionality of the all-optical solution space provided by multi-layer diffractive networks expands linearly as a function of the number of trainable surfaces, $K$, until it reaches the limit defined by the target input and output fields-of-view, i.e., $\min\left(N_{FOV}^2, \left[\sum_{k=1}^{K} N_{Lk}\right] - (K-1)\right)$, as depicted in Equation 7 and Fig. 2. To numerically validate these conclusions, we adopted a deep-learning-based training strategy to design diffractive image classification systems for two distinct datasets (Figs. 3-6) and investigated their performance in terms of blind inference accuracy, learning and generalization performance, classification efficiency and optical signal contrast, confirming the depth advantages provided by multiple diffractive surfaces compared to a single diffractive layer.

These results and conclusions, along with the underlying analyses, broadly cover various types of diffractive surfaces, including, e.g., metamaterials/metasurfaces, nanoantenna arrays, plasmonics and flat-optics-based designer surfaces. We believe that the deeply sub-wavelength design features of, e.g., diffractive metasurfaces can open up new avenues in the design of coherent optical processers by enabling independent control over the amplitude and phase modulation of neurons of a diffractive layer, also providing unique opportunities to engineer the material dispersion properties as needed for a given computational task.

## 4. Materials and Methods

### 4.1. Coefficient and basis vector generation for an optical network formed by two diffractive surfaces

Here, we present the details of the coefficient and basis vector generation algorithm for a network having two diffractive surfaces with the neurons $(N_{L1}, N_{L2})$ to show that it is capable of forming a vectorized transformation matrix in an $N_{L1}+N_{L2}-1$ dimensional subspace of an $N_{FOV}^2$-dimensional complex-valued vector space. The algorithm depends on the consumption of the transmittance values from the first or the second diffractive layer, i.e., $\boldsymbol{T_1}$ or $\boldsymbol{T_2}$, at each step after its initialization. A random neuron is first chosen from $\boldsymbol{T_1}$ or $\boldsymbol{T_2}$, and then a new basis vector is formed. The chosen neuron becomes the coefficient of this new basis vector, which is generated by using the previously chosen transmittance values and appropriate vectors from $\boldsymbol{h_{ij}}$ (Equation 5). The algorithm continues until all the transmittance values are assigned to an arbitrary complex-valued coefficient and uses all the vectors of $\boldsymbol{h_{ij}}$ in forming the basis vectors.

In Table 1, a pseudo-code of the algorithm is also presented. In this table, $C_{1,k}$ and $C_{2,k}$ represent the sets of transmittance values that include $t_{1,i}$ and $t_{2,j}$, which were not chosen before (at time step *k*), from the first and second diffractive surfaces, respectively. Additionally, $c_k = t_{1,i}$ in Step 7 and $c_k = t_{2,j}$ in Step 10 are the complex-valued coefficients that can be independently determined. Similarly, $\boldsymbol{b_k} = \sum_{t_{2,j} \notin C_{2,k}} t_{2,j} \boldsymbol{h_{ij}}$ and $\boldsymbol{b_k} = \sum_{t_{1,i} \notin C_{1,k}} t_{1,i} \boldsymbol{h_{ij}}$ are the basis vectors generated at each step, where $t_{1,i} \notin C_{1,k}$ and $t_{2,j} \notin C_{2,k}$ represent the sets of coefficients that are chosen before. The basis vectors in Step 7 and Step 10 are formed through the linear combinations of the corresponding $\boldsymbol{h_{ij}}$ vectors.

By examining the algorithm in Table 1, it is straightforward to show that the total number of generated basis vectors is $N_{L1}+N_{L2}-1$. That is, at each time step *k*, only one coefficient either from the first or the second layer is chosen, and only one basis vector is created. Since there are $N_{L1}+N_{L2}$-many transmittance values where two of them are chosen together in Step 1, the total number of time steps (coefficient and basis vectors) becomes $N_{L1}+N_{L2}-1$. On the other hand, showing that all the $N_{L1}N_{L2}$-many $\boldsymbol{h_{ij}}$ vectors are used in the algorithm requires further analysis. Without loss of generality, let $\boldsymbol{T_1}$ be chosen $n_1$ times starting from the time step $k = 2$, and then $\boldsymbol{T_2}$ is chosen $n_2$ times. Similarly, $\boldsymbol{T_1}$ and $\boldsymbol{T_2}$ are chosen $n_3$ and $n_4$ times in the following cycles, respectively. This pattern continues until all $N_{L1}+N_{L2}$-many transmittance values are consumed. Here, we show the partition of the selection of the transmittance values from $\boldsymbol{T_1}$ and $\boldsymbol{T_2}$ for each time step *k* into *s* many chunks, i.e.,

$$k = \left\{ \underbrace{2,3,\ldots}_{n_1}, \underbrace{\ldots}_{n_2}, \underbrace{\ldots}_{n_3}, \underbrace{\ldots}_{n_4}, \ \ldots, \underbrace{\ldots N_{L1}+N_{L2}-2, N_{L1}+N_{L2}-1}_{n_s} \right\}$$ 8

To show that $N_{L1}N_{L2}$-many $\boldsymbol{h_{ij}}$ vectors are used in the algorithm regardless of the values of *s* and $n_i$, we first define

$$p_i = n_i + p_{i-2} \text{ for even values of i} \geq 2$$
$$q_i = n_i + q_{i-2} \text{ for odd values of i} \geq 1$$

where $p_0 = 0$ and $q_{-1} = 1$. Based on this, the total number of consumed basis vectors inside each summation in Table 1 (Steps 7 and 10) can be written as:

$$\begin{aligned} n_h = 1 + \sum_{k=2}^{q_1} 1 + \sum_{k=q_1+1}^{p_2+q_1} q_1 + \sum_{k=p_2+q_1+1}^{q_3+p_2} (p_2+1) + \sum_{k=q_3+p_2+1}^{p_4+q_3} q_3 \\ + \sum_{k=p_4+q_3+1}^{q_5+p_4} (p_4+1) + \sum_{k=q_5+p_4+1}^{p_6+q_5} q_5 + \sum_{k=p_6+q_5+1}^{q_7+p_6} (p_6+1) \\ + \quad \cdots \\ + \sum_{k=p_{s-2}+q_{s-3}+1}^{N_{L1}+p_{s-2}} (p_{s-2}+1) + \sum_{k=N_{L1}+p_{s-2}+1}^{N_{L1}+N_{L2}-1} N_{L1} \end{aligned} \quad 9$$

where each summation gives the number of consumed $\boldsymbol{h_{ij}}$ vectors in the corresponding chunk. Please note that based on the partition given by Equation 8, $q_{s-1}$ and $p_s$ become equal to $N_{L1}$ and $N_{L2} - 1$, respectively. One can show, by carrying out this summation, that all the terms except $N_{L1}N_{L2}$ cancel each other out, and therefore, $n_h = N_{L1}N_{L2}$, demonstrating that all the $N_{L1}N_{L2}$-many $\boldsymbol{h_{ij}}$ vectors are used in the algorithm. Here, we assumed that the transmittance values from the first diffractive layer are consumed first. However, even if it were assumed that the transmittance values from the second diffractive layer are consumed first, the result does not change (also see Supplementary Information Section S4.2 and Figure S2).

The Supplementary Information and Table S1 also report an independent analysis of the special case for $N_{L1} = N_{L2} = N_i = N_o = N$ and Table S3 reports the special case of $N_{L2} = N_i = N_o = N$ and $N_{L1} = (K-1)N - (K-2)$, all of which confirm the conclusions reported here. The Supplementary Information also includes an analysis of the coefficient and basis vector generation algorithm for a network formed by three diffractive surfaces ($K$=3) when $N_{L1} = N_{L2} = N_{L3} = N_i = N_o = N$ (see Table S2); also see Supplementary Figure S3 for additional numerical analysis of $K$ = 3 case, further confirming the same conclusions.

### 4.2. Optical forward model

In a coherent optical processor composed of diffractive surfaces, the optical transformation between a given pair of input/output fields-of-view is established through the modulation of light by a series of diffractive surfaces, which we modelled as two-dimensional, thin, multiplicative elements. According to our formulation, the complex-valued transmittance of a diffractive surface, $k$, is defined as:

$$t(x, y, z_k) = a(x, y)\exp\big(j2\pi\phi(x, y)\big) \quad 10$$

where $a(x, y)$ and $\phi(x, y)$ denote the trainable amplitude and the phase modulation functions of diffractive layer $k$. The values of $a(x, y)$, in general, lie in the interval (0, 1), i.e., there is no optical gain over these surfaces, and the dynamic range of the phase modulation is between (0, $2\pi$). In the case of phase-only modulation restriction, however, $a(x, y)$ is kept as 1 (non-trainable) for all the neurons. The parameter $z_k$ defines the axial location of the diffractive layer $k$ between the input field-of-view at $z = 0$ and the output plane. Based on these assumptions, the Rayleigh-Sommerfeld formulation expresses the light diffraction by modelling each diffractive unit on layer $k$ at $(x_q, y_q, z_k)$ as the source of a secondary wave:

$$w_q^k(x, y, z) = \frac{z - z_k}{r^2}\left(\frac{1}{2\pi r} + \frac{1}{j\lambda}\right)\exp\left(\frac{j2\pi r}{\lambda}\right) \quad 11$$

where $r = \sqrt{\left(x - x_q\right)^2 + \left(y - x_q\right)^2 + (z - z_k)^2}$. Combining Equations 10 and 11, we can write the light field exiting the $q^{th}$ diffractive unit of layer $k+1$ as:

$$u_q^{k+1}(x, y, z) = t\left(x_q, y_q, z_{k+1}\right)w_q^{k+1}(x, y, z)\sum_{p\,\epsilon\,S_k} u_p^k\left(x_q, y_q, z_{k+1}\right) \quad 12$$

where $S_k$ denotes the set of diffractive units of layer $k$. From Equation 12, the complex wave field at the output plane can be written as:

$$u^{K+1}(x, y, z) = \sum_{q\,\in\,S_K}\left[t\left(x_q, y_q, z_K\right)w_q^K(x, y, z)\sum_{p\,\epsilon\,S_{K-1}} u_p^{K-1}\left(x_q, y_q, z_K\right)\right] \quad 13$$

where the optical field immediately after the object is assumed to be $u^0(x, y, z)$. In Equation 13, $S_K$ and $S_{K-1}$ denote the set of features at the $K^{th}$ and $(K\text{-}1)^{th}$ diffractive layers, respectively.

### 4.3. Image classification datasets and diffractive network parameters

There are a total of nine image classes in the dataset defined in Fig. 3, corresponding to nine different sets of coordinates inside the input field-of-view, which covers a region of 80λ × 80λ. Each point source lies inside a region of λ × λ, resulting in 6.4K coordinates, divided into nine image classes. Nine classification detectors were placed at the output plane, each representing a data class, as depicted in Fig. 3b. The sensitive area of each detector was set to 25λ × 25λ. In this design, the classification decision was made based on the *maximum* of the optical signal collected by these nine detectors. According to our system architecture, the image in the field-of-view and the class detectors at the output plane were connected through diffractive surfaces of size 100λ × 100λ, and for the multi-layer ($K > 1$) configurations, the axial distance, $d$, between two successive diffractive surfaces was taken as 40λ. With a neuron size of λ/2, we obtained N = 40K (200×200), $N_i$ = 25.6K (160×160) and $N_o$ = 22.5K (9×50×50).

For the classification of the CIFAR-10 image dataset, the size of the diffractive surfaces was taken to be approximately 106.6λ × 106.6λ, and the edge length of the input field-of-view containing the input image was set to be ~53.3λ in both lateral directions. Unlike the amplitude encoded images of the previous dataset (Fig. 3), the information of the CIFAR-10 images was encoded in the phase channel of the input field, i.e., a given input image was assumed to define a phase-only object with the grey levels corresponding to the delays experienced by the incident wavefront within the range [0, λ). To form the phase-only object inputs based on the CIFAR-10 dataset, we converted the RGB samples to greyscale by computing their YCrCb representations. Then, unsigned 8-bit integer values in the Y channel were converted into float32 values and normalized to the range [0, 1]. These normalized greyscale images were then mapped to phase values between [0, 2π). The original CIFAR-10 dataset[49] has 50K training and 10K test images. In the diffractive optical network designs presented here, we used all 50K and 10K images during the training and testing stages, respectively. Therefore, the blind classification accuracy, efficiency and optical signal contrast values depicted in Fig. 6 were computed over the entire 10K test set. Supplementary Figures S4 and S5 demonstrate 600 examples of the greyscale CIFAR-10 images used in the training and testing phases of the presented diffractive network models, respectively.

The responsivity of the 10 class detectors placed at the output plane (each representing one CIFAR-10 data class, e.g., automobile, ship, truck, etc.) was assumed to be identical and uniform over an area of 6.4λ × 6.4λ. The axial distance between two successive diffractive surfaces in the design was assumed to be 40λ. Similarly, the input and output fields-of-view were placed 40λ away from the first and last diffractive layers, respectively.

### 4.4. Loss functions and training details

For a given dataset with C classes, one way of designing an all-optical diffractive classification network is to place C class detectors at the output plane, establishing a one-to-one correspondence between data classes and the opto-electronic detectors. Accordingly, the training of these systems aims to find/optimize the diffractive surfaces that can route most of the input photons, thus the optical signal power, to the corresponding detector representing the data class of a given input object.

The first loss function that we used for the training of diffractive optical networks is the cross-entropy loss, which is frequently used in machine learning for multi-class image classification. This loss function acts on the optical intensities collected by the class detectors at the output plane and is defined as:

$$\mathcal{L} = -\sum_{c \in C} g_c \log(\sigma_c) \quad (14)$$

where $g_c$ and $\sigma_c$ denote the entry in the one-hot label vector and the class score of class c, respectively. The class score $\sigma_c$, on the other hand, is defined as a function of the normalized optical signals, $\boldsymbol{I}'$;

$$\sigma_c = \frac{\exp(I'_c)}{\sum_{c \in C} \exp(I'_c)} \quad (15)$$

Equation 15 is the well-known softmax function. The normalized optical signals $\boldsymbol{I}'$ are defined as $\frac{\boldsymbol{I}}{\max\{\boldsymbol{I}\}} \times T$, where $\boldsymbol{I}$ is the vector of the detected optical signals for each class detector and $T$ is a constant parameter that induces a virtual contrast, helping to increase the efficacy of training.

Alternatively, the all-optical classification design achieved using a diffractive network can be cast as a coherent image projection problem by defining a ground-truth spatial intensity profile at the output plane for each data class and an associated loss function that acts over the synthesized optical signals at the output plane. Accordingly, the mean-squared-error (MSE) loss function used in Fig. 6 computes the difference between a ground-truth intensity profile, $I_g^c(x,y)$, devised for class c and the intensity of the complex wave field at the output plane, i.e., $|u^{K+1}(x,y)|^2$. We defined $I_g^c(x,y)$ as:

$$I_g^c(x,y) = \begin{cases} 1 & if\ x \in D_x^c\ and\ y \in D_y^c \\ 0 & otherwise \end{cases} \quad (16)$$

where $D_x^c$ and $D_y^c$ represent the sensitive/active area of the class detector corresponding to class c. The related MSE loss function, $\mathcal{L}_{mse}$, can then be defined as:

$$\mathcal{L}_{mse} = \int\int \left| |u^{K+1}(x,y)|^2 - I_g^c(x,y) \right|^2 dxdy \quad (17)$$

All network models used in this work were trained using Python (v3.6.5) and TensorFlow (v1.15.0, Google Inc.). We selected the Adam[50] optimizer during the training of all the models, and its parameters were taken as the default values used in TensorFlow and kept identical in each model. The learning rate of the diffractive optical networks was set to 0.001.

# Figures and Tables

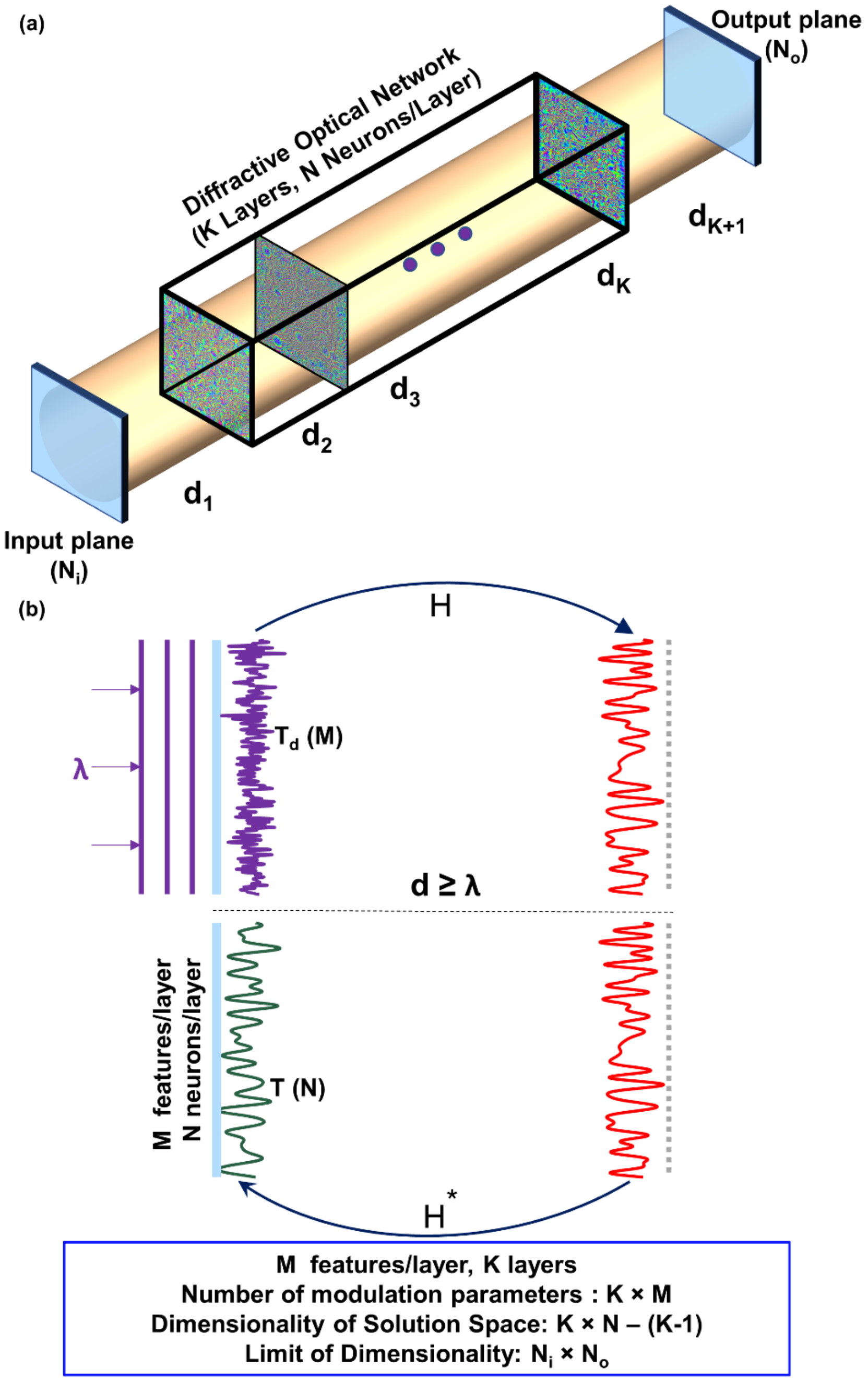

**Fig. 1: Schematic of a multi-surface diffractive network. a** Schematic of a diffractive optical network that connects an input field-of-view (aperture) composed of $N_i$ points to a desired region-of-interest at the output plane/aperture covering $N_o$ points, through $K$ diffractive surfaces with $N$ neurons per surface, sampled at a period of $\lambda/2n$, where $\lambda$ and $n$ represent the illumination wavelength and the refractive index of the medium between the surfaces, respectively. Without loss of generality, $n = 1$ was assumed in this manuscript. **b** The communication between two successive diffractive surfaces occurs through propagating waves when the axial separation ($d$) between these layers is larger than $\lambda$. Even if the diffractive surface has deeply sub-wavelength structures, as in the case of, e.g., metasurfaces, with a much smaller sampling period compared to $\lambda/2$ and many more degrees of freedom ($M$) compared to $N$, the information processing capability of a diffractive surface within a network is limited to propagating modes since $d \geq \lambda$; this limits the effective number of neurons per layer to $N$, even for a surface with $M >> N$. $H$ and $H$* refer to the forward and backward wave propagation, respectively.

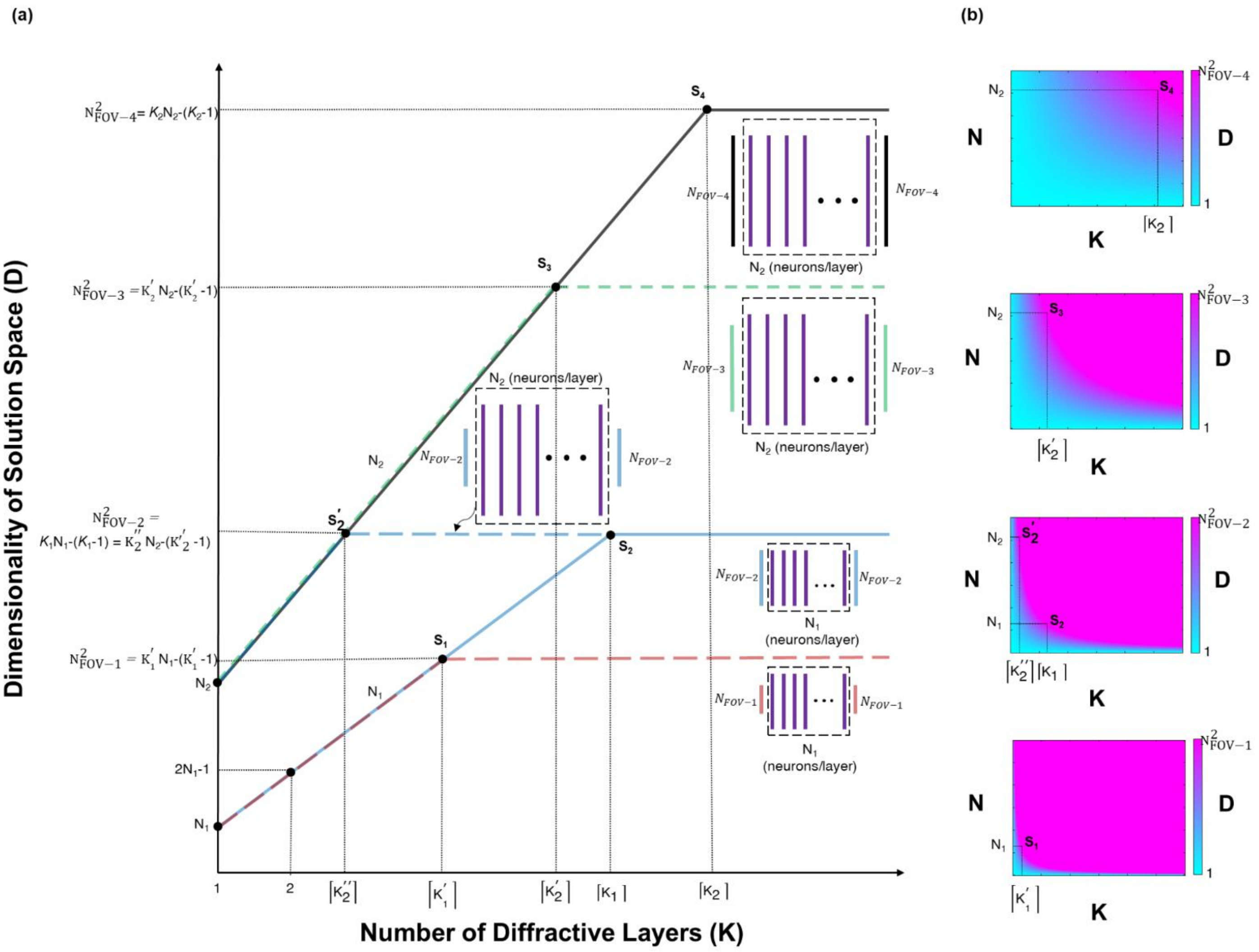

**Fig. 2: Dimensionality ($D$) of the all-optical solution space covered by multi-layer diffractive networks**. **a** The behaviour of the dimensionality of the all-optical solution space as the number of layers increases for two different diffractive surface designs with $N = N_1$ and $N = N_2$ neurons per surface, where $N_2 > N_1$. The smallest number of diffractive surfaces, $\lceil K_S \rceil$, satisfying the condition $K_S N - (K_S - 1) \geq N_i \times N_o$ determines the ideal depth of the network for a given $N$, $N_i$ and $N_O$. For the sake of simplicity, we assumed $N_i = N_o = N_{FOV-i}$, where 4 different input/output fields-of-view are illustrated in the plot, i.e., $N_{FOV-4} > N_{FOV-3} > N_{FOV-2} > N_{FOV-1}$. $\lceil K_S \rceil$ refers to the ceiling function, defining the number of diffractive surfaces within an optical network design. **b** The distribution of the dimensionality of the all-optical solution space as a function of $N$ and $K$ for 4 different fields-of-view, $N_{FOV-i}$, and the corresponding turning points, $S_i$, which are shown in **(a)**. For $K$=1, $d_1 \neq d_2$ is assumed. Also see Supplementary Figures S1-S3 for some examples of $K$=1, 2 and 3.

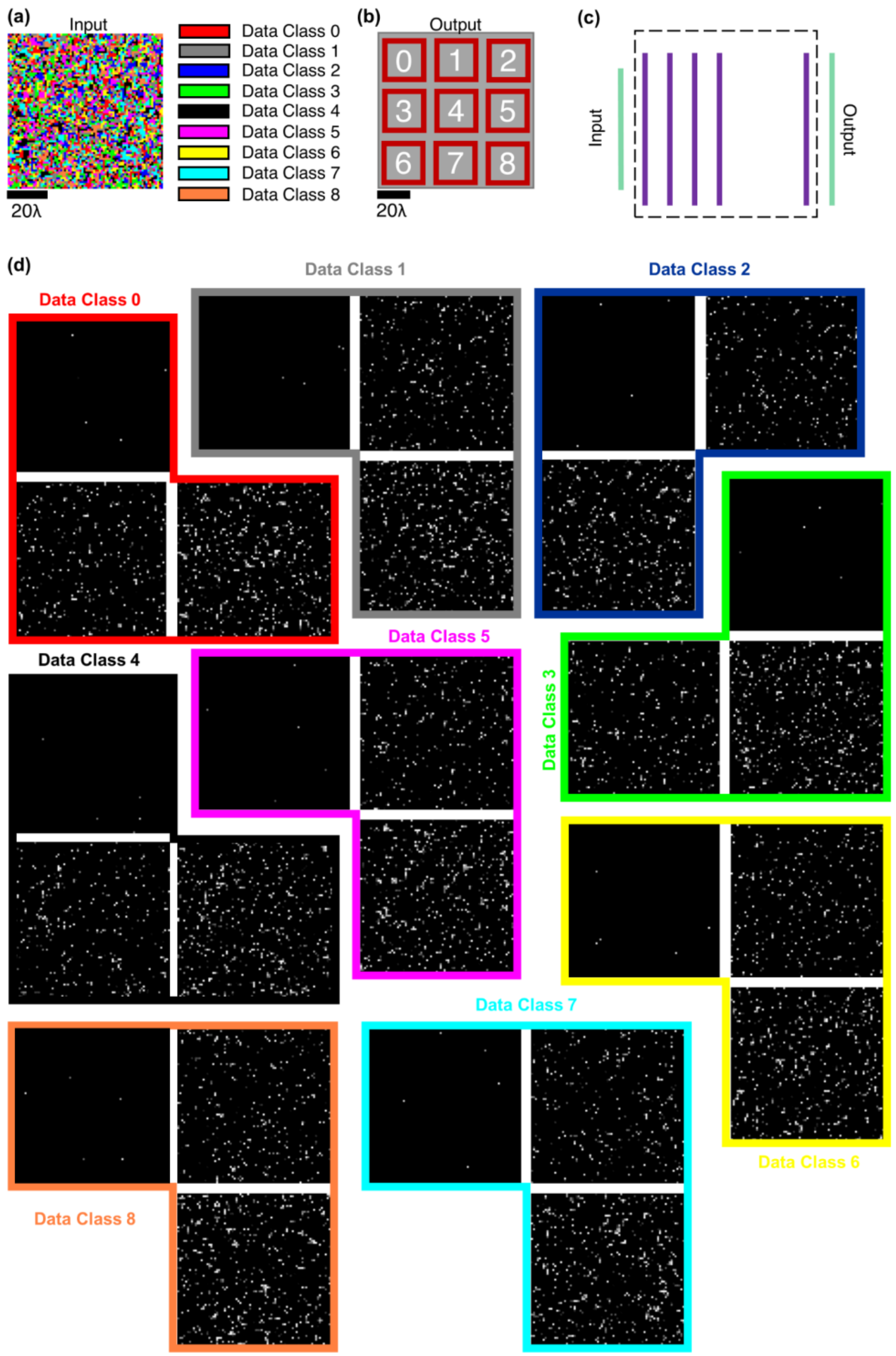

**Fig. 3: Spatially encoded image classification dataset. a** Nine image data classes are shown (presented in different colours), defined inside the input field-of-view (80λ × 80λ). Each λ × λ area inside the field-of-view is randomly assigned to one image data class. An image belongs to a given data class if and only if all of its nonzero entries belong to the pixels that are assigned to that particular data class. **b** The layout of the 9 class detectors positioned at the output plane. Each detector has an active area of 25λ × 25λ, and for a given input image, the decision on class assignment is made based on the *maximum* optical signal among these 9 detectors. **c** Side view of the schematic of the diffractive network layers, as well as the input and output fields-of-view. **d** Example images for 9 different data classes. Three samples for each image data class are illustrated here, randomly drawn from the 3 test datasets ($Te_1$, $Te_{50}$, and $Te_{90}$) that were used to quantify the blind inference accuracies of our diffractive network models (see Fig. 4).

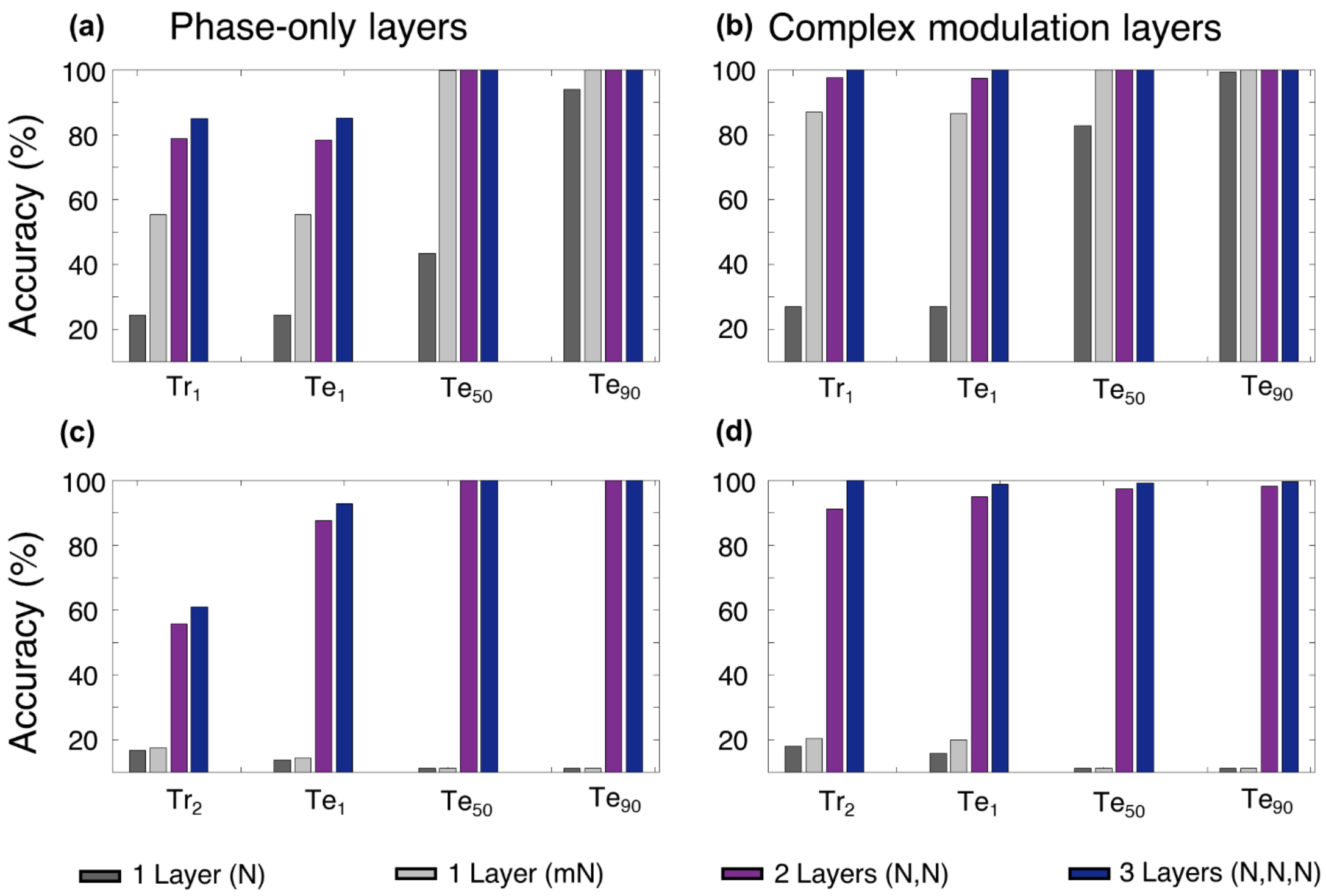

**Fig. 4: Training and testing accuracy results for the diffractive surfaces that perform image classification (Figure 3).** **a** The training and testing classification accuracies achieved by optical network designs composed of diffractive surfaces that control only the phase of the incoming waves; the training image set is $Tr_1$ (200K images). **b** The training and testing classification accuracies achieved by optical network designs composed of diffractive surfaces that can control both the phase and amplitude of the incoming waves; the training image set is $Tr_1$. **c,d** Same as in (**a**,**b**), respectively, except that the training image set is $Tr_2$ (6.4K images). $N$ = 40K neurons, and $mN$ = 122.5K neurons, i.e., $m$>3.

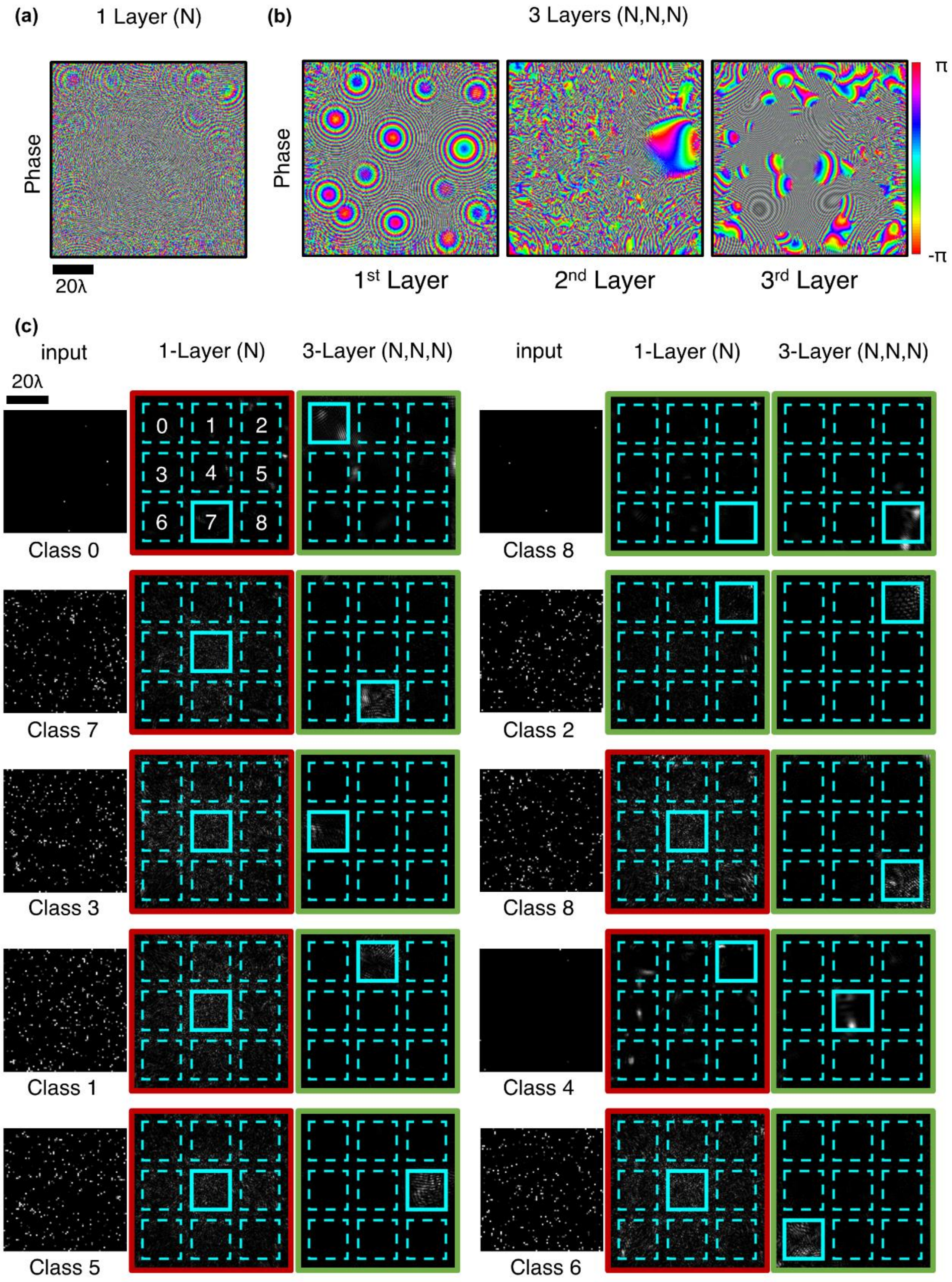

**Fig. 5**: **1- and 3-layer phase-only diffractive network designs and their input-output intensity profiles. a** The phase profile of a single diffractive surface trained with $Tr_1$. **b** Same as in **(a)**, except that there are 3 diffractive surfaces trained in the network design. **c** The output intensity distributions for the 1- and 3-layer diffractive networks shown in **(a)** and **(b)**, respectively, for different input images, which were randomly selected from $Te_1$ and $Te_{50}$. A red (green) frame around the output intensity distribution indicates incorrect (correct) optical inference by the corresponding network. $N$ = 40K.

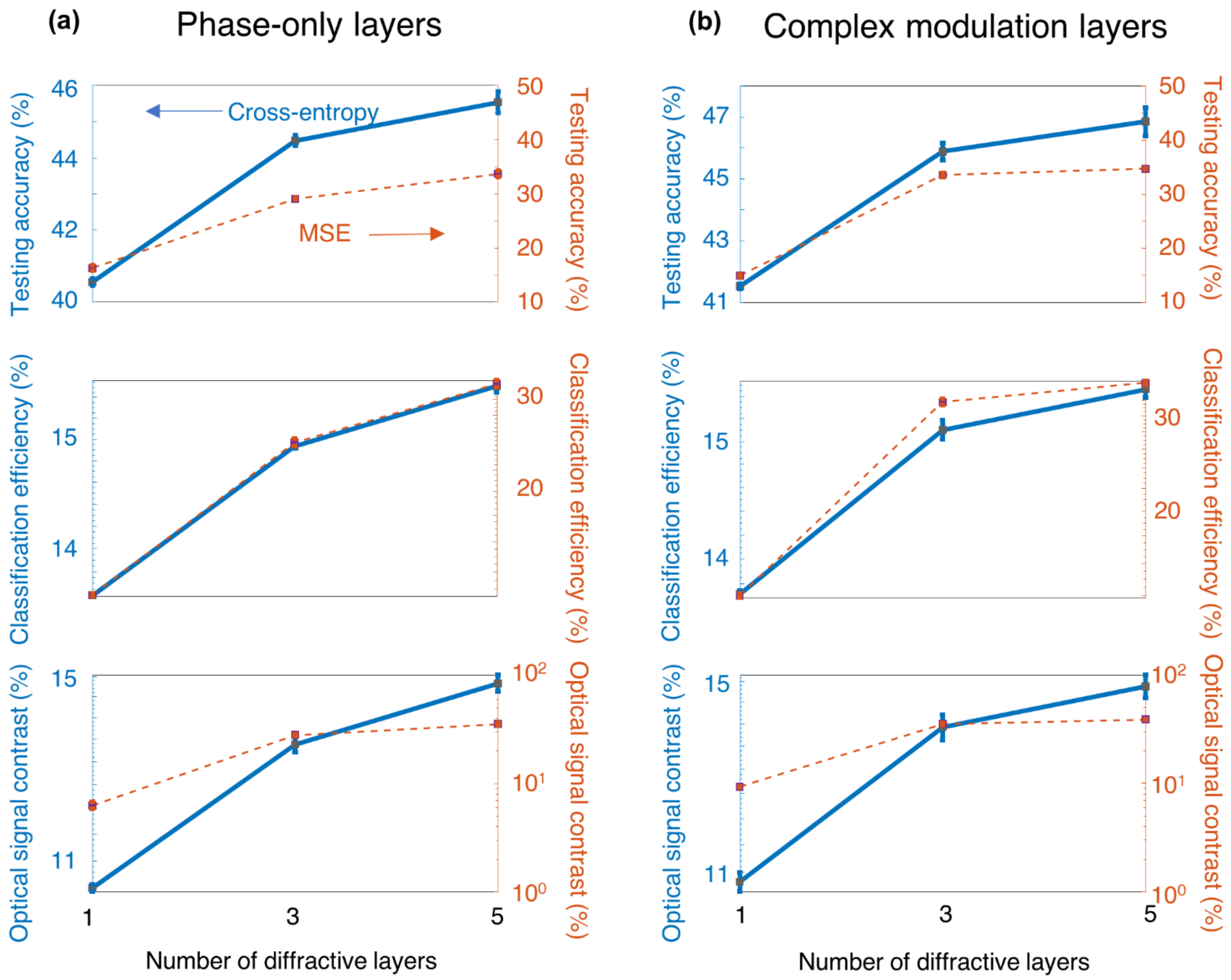

**Fig. 6: Comparison of the 1-, 3- and 5-layer diffractive networks trained for CIFAR-10 image classification using the MSE and cross-entropy loss functions. a** Results for diffractive surfaces that modulate only the phase information of the incoming wave. **b** Results for diffractive surfaces that modulate both the phase and amplitude information of the incoming wave. The increase in the dimensionality of the all-optical solution space with additional diffractive surfaces of a network brings significant advantages in terms of generalization, blind testing accuracy, classification efficiency and optical signal contrast. The classification efficiency denotes the ratio of the optical power detected by the correct class detector with respect to the total detected optical power by all the class detectors at the output plane. Optical signal contrast refers to the normalized difference between the optical signals measured by the ground-truth (correct) detector and its strongest competitor detector at the output plane.

| | |
|---|---|
| 1 | *Randomly choose $t_{1,i}$ from the set $C_{1,1}$ and $t_{2,j}$ from the set $C_{2,1}$, and assign desired values to the chosen $t_{1,i}$ and $t_{2,j}$* |
| 2 | $c_1\boldsymbol{b_1} = t_{1,i}t_{2,j}\boldsymbol{h_{ij}}$ |
| 3 | *k=2* |
| 4 | *Randomly choose $\boldsymbol{T_1}$ or $\boldsymbol{T_2}$ if $C_{1,k} \neq \emptyset$ and $C_{2,k} \neq \emptyset$*<br>*Choose $\boldsymbol{T_1}$ if $C_{1,k} \neq \emptyset$ and $C_{2,k} = \emptyset$*<br>*Choose $\boldsymbol{T_2}$ if $C_{1,k} = \emptyset$ and $C_{2,k} \neq \emptyset$* |
| 5 | *If $\boldsymbol{T_1}$ is chosen in Step 4:* |
| 6 | *Randomly choose $t_{1,i}$ from the set $C_{1,k}$, and assign a desired value to the chosen $t_{1,i}$* |
| 7 | $c_k\boldsymbol{b_k} = t_{1,i}\left(\sum_{t_{2,j} \notin C_{2,k}} t_{2,j}\boldsymbol{h_{ij}}\right)$ |
| 8 | *else:* |
| 9 | *Randomly choose $t_{2,j}$ from the set $C_{2,k}$, and assign a desired value to the chosen $t_{2,j}$* |
| 10 | $c_k\boldsymbol{b_k} = t_{2,j}\left(\sum_{t_{1,i} \notin C_{1,k}} t_{1,i}\boldsymbol{h_{ij}}\right)$ |
| 11 | *k = k+1* |
| 12 | *If $C_{1,k} \neq \emptyset$ or $C_{2,k} \neq \emptyset$:* |
| 13 | *Return to Step 4* |
| 14 | *else:* |
| 15 | *Exit* |

**Table 1.** Coefficient ($c_k$) and basis vector ($\boldsymbol{b_k}$) generation algorithm pseudo-code for an optical network that has two diffractive surfaces. See the theoretical analysis and Equation 6 of the main text. See also Supplementary Tables S1-S3.